\documentclass[aip,showpacs,amsmath,amssymb,amsfonts,lengthcheck,longbibliography,superscriptaddress]{revtex4-1}

\usepackage{changes}

\definechangesauthor[name={AS}, color=magenta]{AS}
\setdeletedmarkup{\color{blue}\sout{#1}}

\usepackage[makeroom]{cancel}

\usepackage{here}
\usepackage{graphicx}
\usepackage{subfigure}
\usepackage{amsthm}
\usepackage{verbatim}
\usepackage{dcolumn}
\usepackage{bm}
\usepackage{epsf}
\usepackage{color}
\usepackage[colorlinks=true,citecolor=blue,linkcolor=blue,urlcolor=blue]{hyperref}%
\usepackage{xcolor}
\usepackage{dsfont}
\newcommand{\bra}[1]{\left\langle #1\right|}
\newcommand{\ket}[1]{\left|#1\right\rangle}

\newcommand{\tr}[1]{\mathrm{tr}\left\{#1\right\}}
\newcommand{\ptr}[2]{\mathrm{tr_{#1}}\left\{#2\right\}}

\newcommand{\lo}[1]{\ln{\left(#1\right)}}
\newcommand{\id}{\mathbb{I}}

\newcommand{\bla}{bla\\bla\\bla\\bla\\bla}

\newcommand{\abs}[2][]{#1| #2 #1|}

\newcommand{\bramatket}[3]{\langle #1 \hspace{1pt} | #2 | \hspace{1pt} #3 \rangle}

\newcommand{\dya}[1]{\ket{#1}\!\bra{#1}}

\newcommand{\ave}[1]{\langle #1\rangle}               
\renewcommand{\geq}{\geqslant}
\renewcommand{\leq}{\leqslant}

\renewcommand{\vec}[1]{\boldsymbol{#1}}  

\newcommand{\ad}{^\dagger}

\newcommand{\eq}{\text{eq}}

{}
{}

\def\P{\mathbb{P}}

\usepackage{amssymb}
\usepackage{dsfont}

\begin{document}

\title{Exchange fluctuation theorems for strongly interacting quantum pumps}
	
\author{Akira Sone}
\affiliation{Department of Physics, University of Massachusetts, Boston, MA 02125, USA}
\email{akira.sone@umb.edu}

\author{Diogo O. Soares-Pinto}
\affiliation{Instituto de F\'isica de S\~{a}o Carlos, Universidade de S\~{a}o Paulo, CP 369, 13560-970, S\~{a}o Carlos, S\~{a}o Paulo, Brazil}
\email{dosp@ifsc.usp.br}

\author{Sebastian Deffner}
\affiliation{Department of Physics, University of Maryland, Baltimore County, Baltimore, MD 21250, USA}
\email{deffner@umbc.edu}

\begin{abstract}
We derive a general quantum exchange fluctuation theorem for multipartite systems with arbitrary coupling strengths by taking into account the informational contribution of the back-action of the quantum measurements, which contributes to the increase in the von-Neumann entropy of the quantum system. The resulting second law of thermodynamics is tighter than the conventional Clausius inequality. The derived bound is the quantum mutual information of the conditional thermal state, which is a thermal state conditioned on the initial energy measurement. These results elucidate the role of quantum correlations in the heat exchange between multiple subsystems.  
\end{abstract}

\maketitle

\section{Introduction}
\label{sec:intro}

The unrivaled success of thermodynamics originates in the very nature of the theory. In contrast to microscopic descriptions, equilibrium thermodynamics formulates universally valid statements based on phenomenological observation. Its central quantity is the entropy, which in a causal universe can never decrease. However, in far from equilibrium situations it is often required to reformulate some statements. This explains the importance and impact of the fluctuation theorems \cite{Evans1993PRL,Gallavotti1995PRL,Jarzynski97,Crooks99,Ortiz2011}, which can be understood as symmetry relations for the probability distribution of entropy production for any physical scenario in the universe.

Remarkably, these fluctuation theorems can be understood as generalizations of the second law of thermodynamics~\cite{jarzynski2013equalities}, and they can be shown to contain previous extensions of thermodynamics away from equilibrium, such as linear response theory~\cite{andrieux2008quantum,andrieux2009fluctuation}. Different versions of these theorems have already been experimentally verified in different systems including biomolecular systems~\cite{collin2005verification}, nuclear magnetic resonance systems~\cite{Batalhao2014}, trapped ion systems~\cite{Huber2008,An15,Smith2018}, and nitrogen-vacancy (NV) centers in diamond~\cite{hernandez2020experimental,hernandez2021experimental}. 

One of the major conceptual building blocks of \emph{stochastic thermodynamics}, a non-equilibrium reformulation of the classical scenario, is to recognize that thermodynamic quantities, such as work, heat \cite{Sekimoto2010} and entropy production \cite{Seifert2005PRL} can be defined along single trajectories. Therefore, it is somewhat natural that the first focus was put on thermodynamic work since it depends only on the action of a controllable external agent \cite{Jarzynski97,Crooks99}. Arguably, heat is a more interesting and complicated quantity, which is related to the uncontrolled forms of the energy exchange. Indeed, fluctuations theorems were also unveiled for heat exchanged between systems \cite{Jarzynski04} as well as more general ``heat-like'' quantities \cite{Hatano2001PRL,Speck2005JPA,Williams2008PRL}. In the present {letter}, we will study the fluctuation theorem for heat, or rather more generally any form of energy, exchanged between interacting quantum systems.

In quantum thermodynamics \cite{DeffnerBook19} the major problem is that classical trajectories become ill-defined. Hence, any thermodynamic quantity defined along a classical trajectory needs to be carefully re-defined. One of the most successful paradigms has been dubbed \emph{two-time measurement} (TTM) scheme ~\cite{Kurchan01, Tasaki00,Talkner07,Campisi11,Deffner2011PRL,Kafri2012,Mazzola2013,Dorner2013,Roncaglia2014,An15,
Deffner2015PRL,Deffner2015PRE,Talkner16,Gardas2016,Bartolotta2018,Gardas2018,Jarzynski04}, in which thermodynamic quantities are determined from two projective measurements of the energy at the beginning and at the end of a process. While such an approach is more geared toward work in isolated quantum systems, it has also found application when considering heat in open quantum systems \cite{Jarzynski04,Deffner2011PRL,Campisi2014JPA,Touil2021PRXQ}.

However, it has been argued that the TTM scheme is thermodynamically inconsistent since the projective energy measurements inevitably destroy quantum coherences~\cite{Marti17}. To overcome this shortcoming of the TTM scheme, recent works proposed alternative paradigms, such as dynamic Bayesian networks~\cite{Micadei20}, the Maggenau-Hill quasiprobability~\cite{Levy19}, and the one-time measurement (OTM) scheme~\cite{Deffner16,Sone20a,Sone21b,Beyer2020,sone2022integrated}. Particularly, in the OTM scheme, the distribution of changes in internal energy is constructed by considering the expectation value of the energy conditioned on the initial energy measurement outcomes. This formalism avoids the second projective measurement and, therefore, the thermodynamic contribution of quantum coherence or the correlations generated by the dynamics  are naturally contained in the formalism \cite{Sone2021entropy}. Therefore, particularly in the strongly coupled systems, the effect of the quantum correlations is non-negligible, so that the OTM scheme naturally becomes the most appropriate paradigm. 

In this letter, we employ the OTM scheme to derive a generalized quantum exchange fluctuation theorem for the multipartite case with arbitrary coupling strengths. As a consequence, we obtain a tighter bound on the net entropy change, which can be characterized by the quantum mutual information of a thermal state conditioned on the initial energy measurement. Our main results elucidate the role of quantum correlations in the heat exchange between arbitrary quantum systems. 

This letter is organized as follows. In Sec.~\ref{sec:notions}, we first summarize the notions and notations which we use in this letter. Then, in Sec.~\ref{sec:ttm}, we discuss the heat distribution for the multipartite systems in the two-time measurement scheme, from which we derive the corresponding quantum exchange fluctuation theorem and the second law of thermodynamics, and recover the bipartite case studied in Ref.~\cite{Jarzynski04}. Third, in Sec.~\ref{sec:otm}, we discuss the heat distribution for the multipartite systems in the one-time measurement scheme, and derive the corresponding quantum exchange fluctuation theorem and second law of thermodynamics with a tighter lower bound, which is associated with the quantum mutual information of the conditional thermal state. Fourth, in Sec.~\ref{sec:ex}, we verify the tighter Clausius bound by employing an example of tripartite XY model. Finally, we conclude in Sec.~\ref{sec:conc}.


\section{Notions and Notations}
\label{sec:notions}

For ease of notation and to avoid clutter in the formulas, we work in units for which the Boltzmann constant $k_B$ and the reduced Planck constant $\hbar$ simply are $k_B=\hbar=1$. We consider a $d$-dimensional multipartite system described by the composite Hilbert space $\mathcal{H}\equiv\bigotimes_{j=1}^n\mathcal{H}_{j}$, where $\mathcal{H}_j$ denotes the Hilbert space of the $j$th subsystem. The corresponding, time-independent Hamiltonians are $H_j$ and $H =\sum_{j=1}^{n} H_j$. Here, $H_j$ is precisely defined as $H_j\equiv H_j\otimes\id_{\overline{j}}$, where $\overline{j}$ denotes the complement of $\mathcal{H}_j$. In the following we will be particularly interested in situations in which the quantum systems are prepared in thermal states at inverse temperatures $\beta_j=1/T_j$. Then the canonical partition functions are $Z_j\equiv \sum_{E_j}e^{-\beta_jE_j}$, where $\{E_j,\ket{E_j}\}$ is the energy eigensystem of the Hamiltonian $H_j$. It will prove convenient to define the energy vectors as $\vec{E}\equiv\left(E_1,\cdots,E_n\right)^T\in\mathbb{R}^n$, energy eigenvectors of $H$ as 
$\ket{\vec{E}}\equiv\ket{E_1,\cdots,E_n}\in\mathbb{C}^{d}$, the inverse temperature vector as $\vec{\beta}\equiv\left(\beta_1,\cdots,\beta_n\right)^T\in\mathbb{R}^n$, and the partition function vector as $\vec{Z}\equiv\left(Z_1,\cdots,Z_n\right)^T\in\mathbb{R}^n$. We also define the product of the partition functions as $\abs{\vec{Z}}\equiv \prod_{j=1}^{n}Z_j$.

\begin{figure}[htp!]
		\centering
		\includegraphics[width=.65\columnwidth]{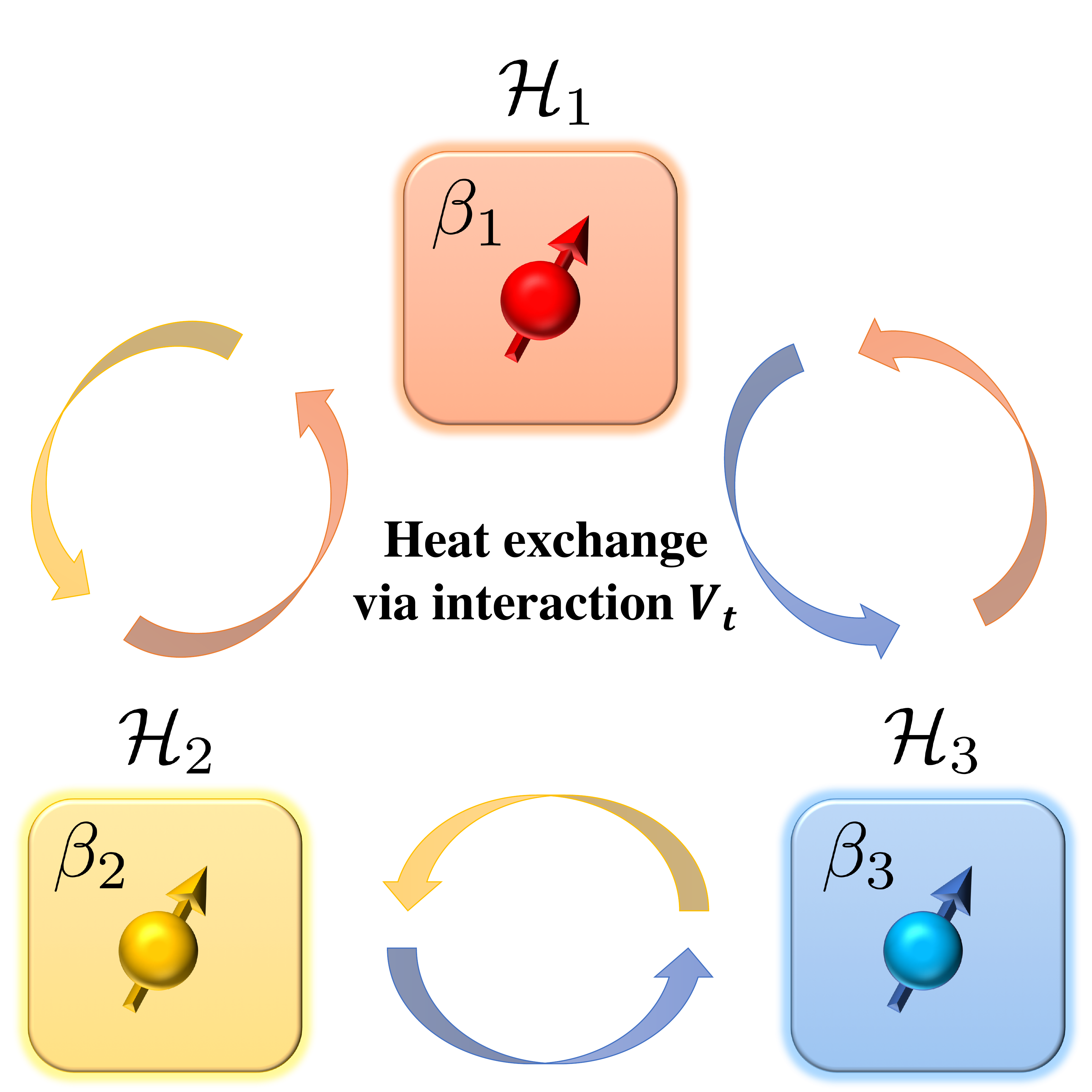}
		\caption{\textbf{Setup}: We illustrate our setup with tripartite systems. The subsystem $\mathcal{H}_1$, $\mathcal{H}_2$ and $\mathcal{H}_3$ are initially prepared in the Gibbs states defined by their inverse temperature and bare Hamiltonians $\{\beta_1,H_1\}$, $\{\beta_2,H_2\}$, and $\{\beta_3,H_3\}$. Then, at time $t=0$, we suddenly turn on the interaction $V_t$ between the tripartite systems, which evolves under the Schr\"{o}dinger's equation. Then, at time $t=\tau$, we suddenly turn off the interaction. The energy is pumped between the subsystems, so that we require the total energy to be conserved.}
\label{fig:setup}
\end{figure}

Consider a situation in which the total system is initially prepared in the product Gibbs state
\begin{equation}
    \rho_0^{\eq} \equiv \bigotimes_{j=1}^{n}\frac{e^{-\beta_j H_j}}{Z_j}=\sum_{\vec{E}}\frac{e^{-\vec{\beta}\cdot\vec{E}}}{\abs{\vec{Z}}}\dya{\vec{E}}\,.
\label{eq:InitialGibbs}
\end{equation}
At the $t=0$ we turn on an interaction $V_t$ and evolve the total system under the unitary operator $U_t$ satisfying the following Schr\"{o}dinger's equation $\partial_t U_t = -i(H+V_t)U_t,~U_0=\id$, where $\partial_t := \partial/\partial t$ (see Fig.~\ref{fig:setup}). 

In the present analysis, we are interested in describing the thermodynamics of the energy that is ``pumped'' between the subsystems $\mathcal{H}_j$, and hence we require the total energy to be conserved, $[U_t,H]=0$. At time $t=\tau$, we switch off the interaction, and as always $\rho_{\tau}\equiv U_{\tau}\rho_0^{\eq}U_{\tau}\ad$.

In complete analogy to Ref.~\cite{Jarzynski04} we define the average heat absorbed by the $j$th system as the energy difference in the system due to the evolution
\begin{equation}
    \ave{Q_j}\equiv \tr{(\rho_{\tau}-\rho_{0}^{\eq})H_j}\,.
\label{eq:AveHeat_j}
\end{equation}
Note that due to energy conservation, we also have $\sum_{j=1}^{n}\ave{Q_j}=0$. It will also be convenient to introduce the vector $\vec{Q}\equiv\left(Q_1,\cdots,Q_n\right)^T\in\mathbb{R}^n$ as a set of the ``stochastic amounts of heat'' in each subsystem (see Fig.~\ref{fig:setup}).

Before we continue, it is important to realize that strictly speaking Eq.~\eqref{eq:AveHeat_j} can be understood as the amounts of exchange thermodynamic heat only as long as the interacting $V_t$ is essentially time-independent. In this case, $V_t$ is suddenly turned on at $t=0$ and suddenly switched off at $t=\tau$. In cases, for which $V_t$ has a more complicated time-dependence, {$\{Q_j\}_{j=1}^{n}$} is strictly speaking the changes of internal energy which may comprise heat and work. In these cases, our scenario describes a quantum energy pump. We stress that the following mathematical analysis remains identical for either case. However, to keep the conceptual arguments as accessible as possible and in close analogy to Ref.~\cite{Jarzynski04} we will be calling $Q_j$ heat.


\section{Two-Time Measurement Scheme}
\label{sec:ttm}

We first discuss the derivation of the quantum exchange fluctuation theorem for multipartite systems in two-time measurement (TTM) scheme. At time $t=0$, we measure the energy of the total system with the Hamiltonian $H$. Suppose we obtain a set of energy values $\vec{E}=\left(E_1,\cdots,E_n\right)^T$, so that the state is projected onto the eigenbases of $H$, $\ket{\vec{E}}\equiv\ket{E_1,\cdots,E_n}$. Then, we evolve the system under the unitary operator $U_{t}$, before we measure $H$ again at $t=\tau$. The outcome of the measurement is $\vec{E}'=\left(E_1',\cdots,E_n'\right)^T$, and the system's state is projected onto $\ket{\vec{E}'}\equiv \ket{E_1',\cdots,E_n'}$. Then, the heat of the $j$th system along the trajectory is simply given by the difference in the energy measurement outcomes $E_j'-E_j$. Following Ref.~\cite{Jarzynski04}, we only focus on the weak coupling regime, where the total energy along this trajectory is approximately conserved $\sum_{j=1}^{n}(E_j'-E_j) \approx 0$. Note that the quantum exchange fluctuation theorem of the TTM scheme also holds for the arbitrary coupling strengths, which can be obtained by applying the characteristic function approach~\cite{esposito2009nonequilibrium}.

Accordingly, the heat distribution becomes
\begin{equation}
\!\!\!\!\!\!P(\vec{Q}) \equiv \sum_{\vec{E},\vec{E}'}
\frac{e^{-\vec{\beta}\cdot\vec{E}}}{\abs{\vec{Z}}}
\abs{\bramatket{\vec{E}'}{U_{\tau}}{\vec{E}}}^2 \, \delta\left(\vec{Q}-(\vec{E}'-\vec{E})\right)\,,
\label{eq:HeatDistTTM}
\end{equation}
where we defined
\begin{equation}
   \delta\left(\vec{Q}-(\vec{E}'-\vec{E})\right)\equiv \prod_{j=1}^{n}\delta\left(Q_j-(E_j'-E_j)\right)\,. 
\end{equation}
Then, the average heat increments are determined from (see Appendix~\ref{app:TTM_Ave_Heat})
\begin{equation}
\ave{Q_j}_{P} = \int d\vec{Q}P(\vec{Q})Q_j =\tr{(\rho_{\tau}-\rho_{0}^{\eq})H_j}\,.
\label{eq:TTM_Ave_Heat}
\end{equation}
It is then only a simple exercise to show that (see Appendix \ref{app:TTMJarzynski})
\begin{equation}
\label{eq:QFT}
\ave{e^{-\vec{\beta}\cdot\vec{Q}}}_{P} = \int d\vec{Q}P(\vec{Q})e^{-\vec{\beta}\cdot\vec{Q}}=1\,,
\end{equation}
where we used the completion relation $\id = \sum_{\vec{E}}\dya{\vec{E}}$ and $\abs{\vec{Z}}=\sum_{\vec{E}'} e^{-\vec{\beta}\cdot\vec{E}'} =\sum_{\vec{E}} e^{-\vec{\beta}\cdot\vec{E}}$. From Jensen's inequality, we immediately also obtain that 
\begin{equation}
    \sum_{j=1}^{n}\beta_j\ave{Q_j}\geq 0\,.
\label{eq:NetEntropyTTM}
\end{equation}

Equations~\eqref{eq:QFT} and \eqref{eq:NetEntropyTTM} readily reduce to the bipartite results of Ref.~\cite{Jarzynski04}. In this case, we recognize $Q_1=-Q_2=Q$ and with $\Delta\beta\equiv\beta_1-\beta_2$ we have
\begin{equation}
    \ave{e^{-\beta_1Q_1-\beta_2Q_2}}_{P}=\ave{e^{-\Delta\beta\cdot Q}}_{P}=1\,,
\label{eq:JarzynskiWojcikBipartite}
\end{equation}
and
\begin{equation}
\Delta\beta\cdot\ave{Q}\geq 0\,.
\label{eq:ClausiusBarpatite}
\end{equation}
More interestingly, we will now derive results equivalent to Eqs.~\eqref{eq:QFT} and \eqref{eq:NetEntropyTTM} within the OTM scheme.


\section{One-Time Measurement Scheme}
\label{sec:otm}

As before, at time $t=0$, we measure the energy of the total system with the total bare Hamiltonian $H$, and then evolve the system under the unitary operator $U_{t}$. In the OTM scheme we do \emph{not} perform the second measurement. Therefore, we define the stochastic heat of a trajectory as the energy difference between the average energy \emph{conditioned} on the initial energy value $\bramatket{\vec{E}}{U_{\tau}\ad H_j U_{\tau}}{\vec{E}}$ and initial energy value $E_j$
\begin{equation}
    \widetilde{Q}_j\equiv \bramatket{\vec{E}}{U_{\tau}\ad H_j U_{\tau}}{\vec{E}}-E_j\,.
\label{eq:NewHeatTrajectory}
\end{equation}
Here, note that $\ket{\vec{E}}$ is the post-measurement state after the first energy measurement; therefore, it is fixed. As before, the total energy is conserved, and we, hence, have $\sum_{j=1}^{n}\widetilde{Q}_j=0$.

Following Ref.~\cite{Deffner16}, the heat distribution $P(\vec{Q})$ is defined as
\begin{equation}
    \widetilde{P}(\vec{Q}) \equiv \sum_{\vec{E}}\frac{e^{-\vec{\beta}\cdot\vec{E}}}{\abs{\vec{Z}}}\,\delta\left(\vec{Q}-\widetilde{\vec{Q}}\right)\,,
\label{eq:NewDistribution}
\end{equation}
where $\widetilde{\vec{Q}}\equiv (\widetilde{Q}_1,\cdots,\widetilde{Q}_n)^T\in\mathbb{R}^n$ and we defined
\begin{equation}
  \delta\left(\vec{Q}-\widetilde{\vec{Q}}\right)\equiv\prod_{j=1}^{n}\delta \left(Q_j-\widetilde{Q}_j\right)\,.
\label{eq:DeltaVector}
\end{equation}
It is easy to see that we have (see Appendix ~\ref{app:OTM_Ave_Heat_j})
\begin{equation}
    \ave{Q_j}_{\widetilde{P}} = \int d\vec{Q}\widetilde{P}(\vec{Q})Q_j =\tr{(\rho_{\tau}-\rho_{0}^{\eq})H_j}\,,
\label{eq:AveConditionalQ}
\end{equation}
which is identical to what we found for the TTM scheme in Eq.~\eqref{eq:AveHeat_j}. 

Deriving the corresponding fluctuation theorem is then again a simple exercise. We obtain
\begin{equation}
    \ave{e^{-\vec{\beta}\cdot\vec{Q}}}_{\widetilde{P}}=\int d\vec{Q}\widetilde{P}(\vec{Q})e^{-\vec{\beta}\cdot\vec{Q}}=\frac{\abs{\widetilde{\vec{Z}}}}{\abs{\vec{Z}}}\,,
\label{eq:JarzynskiPartition}
\end{equation}
where
\begin{equation}
\abs{\widetilde{\vec{Z}}}\equiv \sum_{\vec{E}}\prod_{j=1}^{n}e^{-\beta_j\bramatket{\vec{E}}{U_{\tau}\ad H_j U_{\tau}}{\vec{E}}} 
\label{eq:Normalization}
\end{equation}
is recognized as the normalization of the conditional thermal state \cite{Sone21b,Sone2021entropy}
\begin{equation}
    \widetilde{\rho}_{\tau}\equiv\sum_{\vec{E}} \frac{e^{-\vec{\beta}\cdot\widetilde{\vec{E}}(\vec{E})}}{\abs{\widetilde{\vec{Z}}}}U_{\tau}\dya{\vec{E}}U_{\tau}\ad\,,
\label{eq:ConditionalThermalState}
\end{equation}
where $\widetilde{\vec{E}}(\vec{E})\equiv(\widetilde{E}_1(\vec{E}),\cdots,\widetilde{E}_n(\vec{E}))^T\in\mathbb{R}^n$ with $\widetilde{E}_j(\vec{E})\equiv\bramatket{\vec{E}}{U_{\tau}\ad H_j U_{\tau}}{\vec{E}}$. We have already shown that the conditional thermal state is a thermal state conditioned on the initial energy measurement outcome $\vec{E}$, which maximizes the von-Neumann entropy given the constraint that the ensemble average of the Hamiltonian is fixed~\cite{Sone20a,sone2023conditional}.

In complete analogy to previous results for quantum work \cite{Deffner16}, the corresponding fluctuation theorem for heat exchange simply becomes (see Appendix~\ref{app:OTM_Exchange_Fluctuation_Theorem})
\begin{equation}
    \ave{e^{-\vec{\beta}\cdot\vec{Q}}}_{\widetilde{P}} = e^{-S(\widetilde{\rho}_{\tau}||\rho_{0}^{\eq})}\,. 
\label{th:main1}
\end{equation}
where $S(\widetilde{\rho}_{\tau}||\rho_{0}^{\eq})=\tr{\widetilde{\rho}_{\tau} \lo{\widetilde{\rho}_{\tau}}}-\tr{\widetilde{\rho}_{\tau} \lo{\rho_{0}^{\eq}}}$ is the quantum relative entropy measuring the ``distance'' of the conditional thermal state to the corresponding Gibbs state. Then, from Jensen's inequality, we also have
\begin{equation}
    \sum_{j=1}^{n}\beta_j\ave{Q_j}\geq S(\widetilde{\rho}_{\tau}||\rho_0^{\eq})\,.
\label{eq:newClausius}
\end{equation}
This is a sharpened statement of the second law, which sets a tighter bound on the net energy exchange. 

For multipartite systems, the quantum mutual information of $\widetilde{\rho}_{\tau}$ is defined as \cite{Watanabe1960}
\begin{equation}
\widetilde{I}_{\tau}(1:\cdots:n) \equiv \sum_{j=1}^{n}S(\widetilde{\rho}_{j,\tau})-S(\widetilde{\rho}_{\tau})\,,
\end{equation}
where $\widetilde{\rho}_{j,\tau}\equiv \ptr{\overline{j}}{\widetilde{\rho}_{\tau}}$ is the reduced state of $\widetilde{\rho}_{\tau}$ of the $j$th system. Writing $\gamma_{\tau}$ as the product state of these reduced states, 
\begin{equation}
\gamma_{\tau}\equiv \bigotimes_{j=1}^{n}\widetilde{\rho}_{j,\tau}\,,
\end{equation}
we have
\begin{equation}
S(\widetilde{\rho}_{\tau}||\rho_{0}^{\eq})=\widetilde{I}_{\tau}(1:\cdots:n)+S\left(\gamma_{\tau}||\rho_{0}^{\eq}\right)\,.
\end{equation}
From Eq.~\eqref{eq:newClausius}, we can obtain our second main result
\begin{equation}
    \sum_{j=1}^{n}\beta_j\ave{Q_j}\geq \widetilde{I}_{\tau}(1:\cdots:n)\,.
\label{cor:main2}
\end{equation}
Thus we have found, that the net heat exchanged in multipartite systems is lower bounded by the amount of mutual information between the subsystems.


\section{Example: Tripartite XY Model}
\label{sec:ex}
\begin{figure}[htp!]
\centering
\includegraphics[width=1\columnwidth]{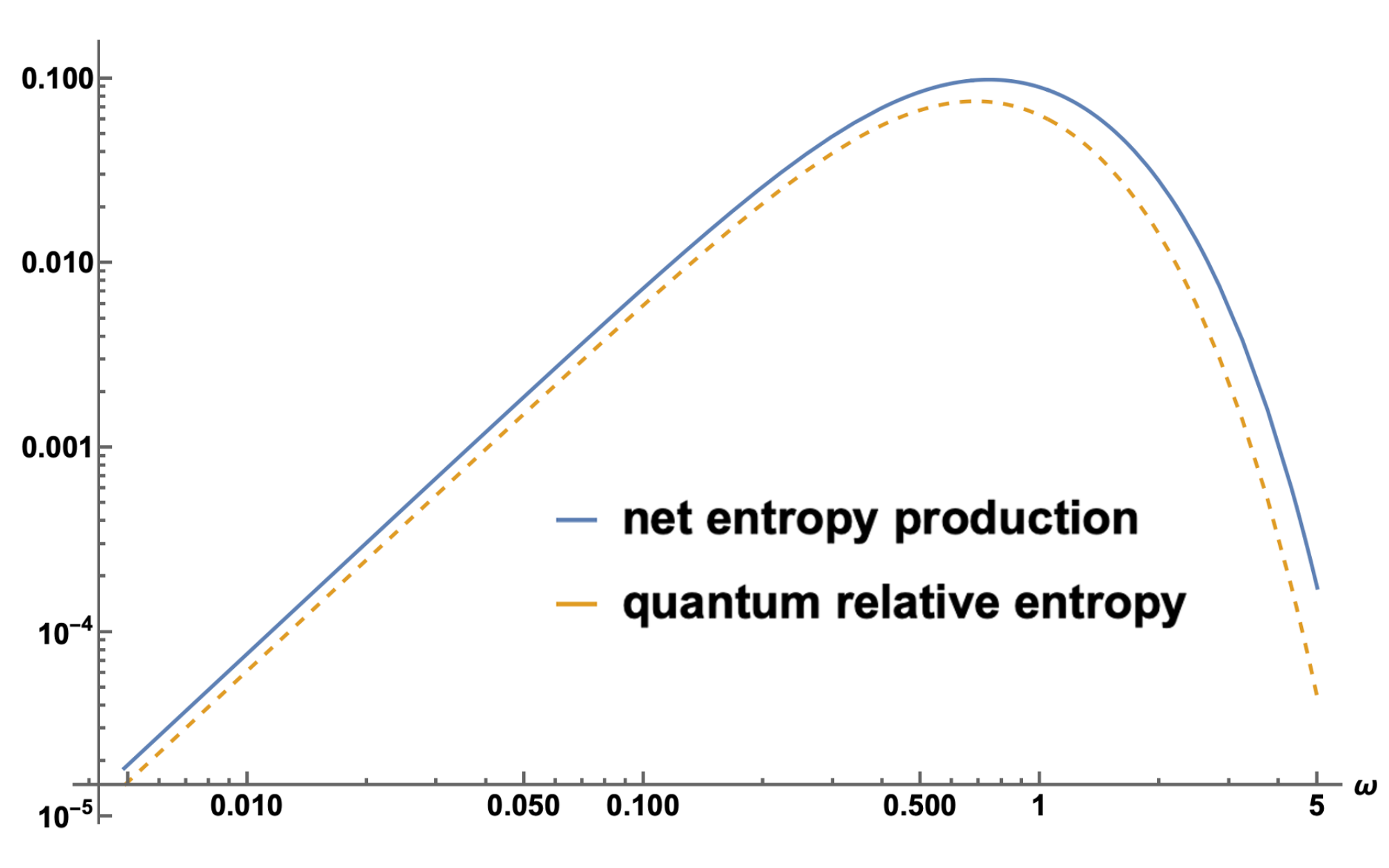}
\caption{\textbf{Three-qubit example}: We plot the net entropy production and the quantum relative entropy as a function of the transverse field $\omega\in[0,~5.0]$ with $\beta_1=1$, $\beta_2=2$, $\beta_3=3$, and $\tau=4.0$. We can verify $\sum_{j=1}^{3}\beta_j\ave{Q_j}\geq S(\widetilde{\rho}_{\tau}||\rho_0^{\eq})$.}
\label{fig:3qubit}
\end{figure}

As an example, we consider a three-qubit system, whose Hamiltonian is given by the $XY$ model
\begin{equation}
    H_t=\omega(\sigma_{z}^{(1)}+\sigma_{z}^{(2)}+\sigma_{z}^{(3)})+V_t\,,
\end{equation}
where we define
\begin{equation}
    \!\!\!\!\!\!V_t \!\equiv\!
    \begin{cases}
    0\!&(t<0,t>\tau)\\
    \sum_{j=1}^{3}\sum_{i<j}(\sigma_x^{(i)}\sigma_x^{(j)}+\sigma_y^{(i)}\sigma_y^{(j)})\!&(0\leq t\leq\tau)
    \end{cases}
\end{equation}
with $\{\id,\sigma_x,\sigma_y,\sigma_z\}$ the Pauli matrices. Here, the bare Hamiltonian for each system is $H_j=\omega\sigma_z^{(j)}$ ($j=1,2,3$). 
Then, we plot the net entropy production $\sum_{j=1}^{3}\beta_j\ave{Q_j}$ and the quantum relative entropy $S(\widetilde{\rho}_{\tau}||\rho_0^{\text{eq}})$ with $\omega\in[0,~5.0]$, $\beta_1=1$, $\beta_2=2$, $\beta_3=3$ and $\tau=4.0$. Then, we obtain the following result (see Fig.~\ref{fig:3qubit}), which verifies Eq.~\eqref{eq:newClausius}.  

\section{Conclusion}
\label{sec:conc}

In conclusion, we have employed the one-time measurement scheme to derive the generalized quantum exchange fluctuation theorem for multipartite systems, which includes the information about the quantum correlations. From it, we have also obtained a tighter lower bound on the net entropy change, which is set by the quantum mutual information of the thermal state conditioned on the initial energy measurement outcomes. These results elucidate the role of quantum correlations in the heat exchange in a generic scenario, where the effect of the quantum correlation due to the interaction between multiple subsystems is non-negligible.

\section*{Acknowledgments}

A.S. gratefully acknowledges startup funding supported by the University of Massachusetts, Boston. D.O.S.P. acknowledges the Brazilian funding agencies CNPq (Grant No. 307028/2019-4), FAPESP (Grant No. 2017/03727-0), and the Brazilian National Institute of Science and Technology of Quantum Information (INCT/IQ). S.D. acknowledges support from  the U.S. National Science Foundation under Grant No. DMR-2010127 and the John Templeton Foundation under Grant No. 62422.

\section*{Author Declarations}
\subsection*{Conflict of interest}
The authors have no conflicts to disclose.

\section*{Data availability statement}
The data that support the findings of this study are available from the corresponding author upon reasonable request.

\appendix

\section{Proof of Eq.~\eqref{eq:TTM_Ave_Heat}}
\label{app:TTM_Ave_Heat}
In this section, we demonstrate the proof of Eq.~\eqref{eq:TTM_Ave_Heat}. From Eq.~\eqref{eq:HeatDistTTM}, we have 
\begin{equation}
\begin{split}
\ave{Q_j}_{P} 
    =& \int d\vec{Q} Q_j P(\vec{Q})\\
    =&\frac{1}{\abs{\vec{Z}}}\sum_{\vec{E}',\vec{E}}(E_j'-E_j) e^{-\vec{\beta}\cdot\vec{E}}\abs{\bramatket{\vec{E}'}{U_{\tau}}{\vec{E}}}^2\,.
\end{split}
\end{equation}
Here, note that precisely, because the bare Hamiltonian is time-independent, $H_j$ is 
\begin{equation}
H_j\equiv \sum_{E_j}E_j\dya{E_j}\otimes \id_{\overline{j}} = \sum_{E_j'}E_j'\dya{E_j'}\otimes \id_{\overline{j}}\,.  
\end{equation}
Also, we have
\begin{equation}
    \sum_{\vec{E}'}\dya{\vec{E}'}=\id\,.
\end{equation}
These lead to
\begin{equation}
    \ave{Q_j}_{P} = \tr{(\rho_{\tau}-\rho_0^{\eq})H_j}=\ave{Q_j}\,.
\end{equation}

\section{Proof of Eq.~\eqref{eq:QFT}}
\label{app:TTMJarzynski}
In this section, we demonstrate the proof of Eq.~\eqref{eq:QFT}. From Eq.~\eqref{eq:HeatDistTTM}, we have 
\begin{equation}
\begin{split}
    \ave{e^{-\vec{\beta}\cdot\vec{Q}}}_{P} =& \int d\vec{Q}e^{-\vec{\beta}\cdot\vec{Q}}P(\vec{Q})\\
    =&
    \sum_{\vec{E},\vec{E}'}\frac{e^{-\vec{\beta}\cdot\vec{E}'}}{\abs{\vec{Z}}}\abs{\bramatket{\vec{E}'}{U_{\tau}}{\vec{E}}}^2\,.
 \end{split}
\end{equation}
Here, we have
\begin{equation}
\sum_{\vec{E}}\dya{\vec{E}}=\id
\end{equation}
and 
\begin{equation}
\abs{\vec{Z}}=\sum_{\vec{E}'}e^{-\vec{\beta}\cdot\vec{E}'} = \sum_{\vec{E}}e^{-\vec{\beta}\cdot\vec{E}}
\end{equation}
because of the time-independence of the bare Hamiltonian. Therefore, we can obtain
\begin{equation}
    \ave{e^{-\vec{\beta}\cdot\vec{Q}}}_P = 1\,.
\end{equation}

\section{Proof of Eq.~\eqref{eq:AveConditionalQ}}
\label{app:OTM_Ave_Heat_j}
In this section, we demonstrate the proof of Eq.~\eqref{eq:AveConditionalQ} in details. From Eqs.~\eqref{eq:NewHeatTrajectory}, \eqref{eq:NewDistribution}, and \eqref{eq:DeltaVector}, the average conditional heat of the $j$-th system is given by 
\begin{equation}
\begin{split}
    \ave{Q_j}_{\widetilde{P}} 
    &= \int d\vec{Q} Q_j\widetilde{P}(\vec{Q})\\
    &=\sum_{\vec{E}}\frac{e^{-\vec{\beta}\cdot\vec{E}}}{\abs{\vec{Z}}}\bramatket{\vec{E}}{U_{\tau}\ad H_j U_{\tau}}{\vec{E}}-\sum_{E_j}\frac{e^{-\beta_jE_j}}{Z_j}E_j\,.
\end{split}
\end{equation}
Here, we have 
\begin{equation}
\begin{split}
\tr{\rho_{\tau}H_j}& = \tr{U_{\tau}\rho_{0}^{\eq}U_{\tau}\ad H_j}\\
&=\sum_{\vec{E}}\frac{e^{-\vec{\beta}\cdot\vec{E}}}{\abs{\vec{Z}}}\bramatket{\vec{E}}{U_{\tau}\ad H_j U_{\tau}}{\vec{E}}\,.
\end{split}
\end{equation}
Also, from Eq.~\eqref{eq:InitialGibbs}, we have 
\begin{equation}
\sum_{E_j}\frac{e^{-\beta_jE_j}}{Z_j}E_j = \tr{\rho_{0}^{\eq}H_j}\,.
\end{equation}
Therefore, we can finally obtain Eq.~\eqref{eq:AveConditionalQ} 
\begin{equation}
    \ave{Q_j}_{\widetilde{P}} = \tr{(\rho_{\tau}-\rho_{0}^{\eq}) H_j}=\ave{Q_j}\,.
\end{equation}

\section{Proof of Eq.~\eqref{th:main1}}
\label{app:OTM_Exchange_Fluctuation_Theorem}

In this section, we demonstrate the proof of Eq.~\eqref{th:main1} in details. From Eq.~\eqref{eq:JarzynskiPartition}, we have 

\begin{equation}
    \ave{e^{-\vec{\beta}\cdot\vec{Q}}}_{\widetilde{P}} = \frac{\abs{\widetilde{\vec{Z}}}}{\abs{\vec{Z}}}\,,
\end{equation}
where $\abs{\vec{Z}}$ is defined in Eq.~\eqref{eq:Normalization}
\begin{align}
    \abs{\widetilde{\vec{Z}}}\equiv \sum_{\vec{E}}\prod_{j=1}^{n}e^{-\beta_j\bramatket{\vec{E}}{U_{\tau}\ad H_j U_{\tau}}{\vec{E}}}\,.
\end{align}
The conditional thermal state is defined in Eq.~\eqref{eq:ConditionalThermalState} as 
\begin{equation}
    \widetilde{\rho}_{\tau}\equiv\sum_{\vec{E}} \frac{e^{-\vec{\beta}\cdot\widetilde{\vec{E}}(\vec{E})}}{\abs{\widetilde{\vec{Z}}}}U_{\tau}\dya{\vec{E}}U_{\tau}\ad\,,
\end{equation}
and let us compute the quantum relative entropy of $\widetilde{\rho}_{\tau}$ with respect to the initial state $\rho_{0}^{\eq}$. The quantum relative entropy is given by 
\begin{equation}
    S(\widetilde{\rho}_{\tau}||\rho_{0}^{\eq}) = -S(\widetilde{\rho}_{\tau})-\tr{\widetilde{\rho}_{\tau}\ln\rho_0^{\eq}}\,.
\end{equation}

First, the von-Neumann entropy of $\widetilde{\rho}_{\tau}$ is given by  
\begin{equation}
\begin{split}
    S(\widetilde{\rho}_{\tau}) & = -\tr{\widetilde{\rho}_{\tau}\ln \widetilde{\rho}_{\tau}}\\
    &=\sum_{\vec{E}}(\vec{\beta}\cdot\widetilde{\vec{E}}(\vec{E}))\tr{\widetilde{\rho}_{\tau}U_{\tau}\dya{\vec{E}}U_{\tau}\ad}+\ln\abs{\widetilde{\vec{Z}}}\\
    &=\sum_{\vec{E}}\sum_{j=1}^{n}\beta_j\bramatket{\vec{E}}{U_{\tau}\ad H_j U_{\tau}}{\vec{E}}\frac{e^{-\vec{\beta}\cdot\widetilde{\vec{E}}(\vec{E})}}{\abs{\widetilde{\vec{Z}}}}+\ln\abs{\widetilde{\vec{Z}}}\\
    &=\sum_{j=1}^{n}\beta_j \tr{\widetilde{\rho}_{\tau}H_j}+\ln\abs{\widetilde{\vec{Z}}}\,.
\end{split}
\end{equation}

Second, the quantum cross entropy $-\tr{\widetilde{\rho}_{\tau}\ln\rho_{0}^{\eq}}$ is given by 
\begin{equation}
\begin{split}
    -\tr{\widetilde{\rho}_{\tau}\ln\rho_{0}^{\eq}}&=-\tr{\widetilde{\rho}_{\tau}\ln\frac{e^{-\sum_{j=1}^{n}\beta_j H_j}}{\abs{\vec{Z}}}}\\
    &=\sum_{j=1}^{n}\beta_j\tr{\widetilde{\rho}_{\tau}H_j}+\ln\abs{\vec{Z}}\,.
\end{split}
\end{equation}

Therefore, we can obtain 
\begin{equation}
    S(\widetilde{\rho}_{\tau}||\rho_{0}^{\eq}) = -\ln\frac{\abs{\widetilde{\vec{Z}}}}{\abs{\vec{Z}}}\,.
\end{equation}
From Eq.~\eqref{eq:JarzynskiPartition}, we can finally obtain Eq.~\eqref{th:main1}
\begin{equation}
    \ave{e^{-\vec{\beta}\cdot\vec{Q}}}_{\widetilde{P}} = e^{-S(\widetilde{\rho}_{\tau}||\rho_{0}^{\eq})}\,.
\end{equation}

\bibliography{quantum_pump}

\end{document}